\documentclass[nolongbibliography,superscriptaddress,showpacs,showkeys,amsmath,amssymb,eqsecnum,aps,prd,nofootinbib]{revtex4-2}

\usepackage[colorlinks, allcolors=Blue]{hyperref}

\usepackage{microtype}
\usepackage{orcidlink}
\usepackage{booktabs}
\usepackage{tensor, mathtools, bm}

\usepackage[dvipsnames]{xcolor}
\usepackage{graphicx}
\usepackage{epsfig}
\usepackage[varbb, varvw]{newtx}

\begin{document}
\title{Quantum Gravity in the Landau background gauge}
\begin{abstract}
    We examine, in the Landau background gauge, the effective action of a quantum theory of gravitational and matter fields. We analyse the BRST invariance of this theory and study the Ward identities which implement the symmetry of the background effective action. The off-shell and on-shell counterterm Lagrangians are explicitly calculated in this gauge at one-loop order. We argue that the matter self-energy in the Landau background gauge, which is finite to this order, may be finite also at higher loops. 
\end{abstract}

\author{J. Frenkel\orcidlink{0000-0003-1178-1001}}
\email{jfrenkel@if.usp.br}
\affiliation{Instituto de F\'{\i}sica, Universidade de S\~ao Paulo, S\~ao Paulo, SP 05508-090, Brazil}

\author{S. Martins-Filho\orcidlink{0000-0003-4195-2713}}
\email{s.martins-filho@unesp.br}
\affiliation{Instituto de F\'{\i}sica Te\'orica, Universidade Estadual Paulista (UNESP), Rua Dr.\ Bento Teobaldo Ferraz 271 - Bloco II, 01140-070 S\~ao Paulo, SP, Brazil}

\date{\today}

\maketitle

\section{Introduction} \label{sec:intro}

In quantum field theory, there are two ways in which a model can be perturbatively renormalizable. The first is a strict renormalizability, when the theory has a finite number of terms in the Lagrangian that cancel out all infinities, which occurs in the Standard Model. The second way, although non-renormalizable in the strict sense, may be thought to be renormalizable in a more general sense, when an infinite number of counterterms are needed to cancel all infinities.
This is the case of Quantum Gravity, where the divergences can be cancelled by introducing all possible counterterms consistent with the symmetries of the theory. The current point of view is that general relativity can be regarded as a low-energy effective field theory. Despite the presence of an infinite number of free parameters,this theory retains a good deal of predictive power that follows in
consequence of the properties of unitarity and analyticity  of the S-matrix. This fact allows us to calculate the non-analytic parts of Feynman amplitudes and can be used to carry out useful perturbative calculations also at higher loop orders~\cite{Weinberg:1978kz, Goroff:1985th, vandeVen:1991gw, Gomis:1995jp, Weinberg:1995mt, Donoghue:2022eay}.

  The use of the  Landau gauge leads to a significant simplification of calculations and is very useful  for the understanding of many physical phenomena. This gauge has been widely used in QED \cite{itzyksonQuantumFieldTheory1980, berestetskii1982}  and  in QCD, where it has been employed in the study of phase transitions through numerical simulation of lattice QCD \cite{Creutz:1983njd}. Moreover, the Landau gauge has many suitable properties for  the analysis of the infrared properties of QCD, which are manifest in the realm of its non-perturbative sector. This study has been developed through approaches such as the functional renormalization group or the Dyson-Schwinger equations and provides  a description of QCD in terms of the quark and gluon correlation functions, which exhibit characteristic mass gaps \cite{Cyrol:2016tym} (see also, for example, the reviews \cite{Berges:2000ew, Binosi:2009qm, Maas:2011se} and references cited therein).

Some of these studies have been done by using the background field method \cite{Grassi:2004yq}, which is convenient because it preserves the gauge symmetry at all stages of the calculations \cite{Abbott:1980hw, Barvinsky:1985an, Barvinsky:1993zg, Kamenshchik:2014waa, Frenkel:2018xup, Brandt:2022und}.
In quantum gravity, the Landau gauge has also been widely employed in the context of the functional renormalization group \cite{Kawasaki:1981zw, Eichhorn:2010tb, Gies:2015tca, Ohta:2016jvw, Ohta:2016npm, Christiansen:2017bsy}, since the Landau gauge corresponds to a fixed point of the renormalization group flow~\cite{Ellwanger:1995qf, Litim:1998qi}.

In this work, we aim to discuss some pertinent aspects of the quantization of an interacting quantum theory of gravitational and scalar fields in the Landau background gauge. Using the Becchi-Rouet-Stora-Tyutin (BRST) symmetry \cite{Becchi:1975nq, Tyutin:1975qk}, we outline in Sec.\ II some basic elements of this theory and discuss the Slavnov-Taylor identity that reflects this symmetry.
In Sec.\ III, we evaluate in this gauge the off-shell counterterm Lagrangian at one-loop order, which turns out to be much simpler than that computed in other background gauges \cite{Frenkel:2026zwc}. 
However, on-shell, one can verify that this Lagrangian yields a result that is consistent with the Kallosh-DeWitt theorem \cite{Kallosh:1974yh, DeWitt:1967ub}, which implies that the on-shell effective action must be independent of the gauge-fixing parameters. 
In Sec.\ IV, we deduce the Ward identities which implement the invariance of the background effective action. Such identities may imply the finiteness, in the Landau background gauge, of the scalar self-energy at higher orders.
A brief discussion of the results is given in Sec.\ V\@. Some useful details of the calculations are presented in the Appendices.

\section{BRST quantization in the Landau background gauge}

Let us consider a theory of gravitational $ g $ and massless scalar $ \phi $ fields described by the Lagrangian \cite{tHooft:1974toh, *thooft1993}
\begin{equation}\label{eq:21}
    \mathcal{L}^{\text{inv}} ( g, \phi ) = - \frac{1}{\kappa^{2}} \sqrt{ g} \left [ {R}  + \frac{\kappa^{2}}{2} g^{\mu \nu} \partial_{\mu} \phi \partial_{\nu} \phi\right ],
\end{equation}
where $ \kappa^{2} = 16 \pi G_{N} $, $G_{N} $ is the Newton constant  and $ {R} $ is the Ricci scalar.
In order to preserve the basic diffeomorphism of the theory, we use the background field method, where the fields are splitted into a background and a quantum part as
\begin{equation}\label{eq:22}
    g_{\mu \nu} = \bar{g}_{\mu \nu} + \kappa h_{\mu \nu} ; \quad \phi = \bar{\phi} + \varphi.
\end{equation}
Here, $ \bar{g}_{\mu \nu} $ and $ \bar{\phi} $ are background fields, while $ h_{\mu \nu} $ and $ \varphi $ are quantum fields.

The Lagrangian \eqref{eq:21}, with the substitution \eqref{eq:22} is invariant under the gauge transformation
\begin{subequations}\label{eq:23}
    \begin{align}\label{eq:23a}
        \delta \bar{g}_{\mu \nu} ={}&0, \quad  \delta \bar{\phi} =0; \\
        \delta h_{\mu \nu} ={}& \mathsf{D}_{\mu} \epsilon_{\nu} + \mathsf{D}_{\nu} \epsilon_{\mu}  ; \quad  \delta \varphi = \kappa \epsilon^{\alpha} \partial_{\alpha} ( \bar{\phi} +  \varphi ).
    \end{align}
\end{subequations}
where $ \mathsf{D}_{\mu} $ is the full covariant derivative.
To quantize the theory in a manner that maintains its background symmetry, one must introduce a gauge-fixing term which breaks only the quantum symmetry. To this end, a convenient choice is the generalised de Donder gauge ($ \xi $ is a generic gauge parameter)
\begin{equation}\label{eq:24}
    \mathcal{L}_{\text{gf}} (\bar{g}, \bar{\phi}, h, \varphi ) = -\frac{1}{2 \xi} \sqrt{\bar{g}} \left( \bar{\mathsf{D}}_{\nu} h^{\mu \nu} - \frac{1}{2} \bar{\mathsf{D}}^{\mu} h_{\alpha}^{\alpha} 
     \right)^{2}
\end{equation}
where $ \bar{\mathsf{D}}_{\mu} $ is the covariant derivative with respect to the background field $ \bar{g}_{\mu \nu} $.
The Landau background gauge is obtained in the limit $\xi \to 0$. After inverting the two point function, one may take in the propagator the limit $ \xi \to 0$. But, the interaction vertices arising from Eq.~\eqref{eq:24} are individually divergent in this limit, so that setting $\xi \to 0$ at the outset would introduce spurious divergences that only cancel once the different contributions to a given calculation are combined. Hence, using Eq.~\eqref{eq:24}, the gauge parameter $\xi$ must be kept finite throughout the intermediate steps, and the limit can only be taken at the end, after these cancellations have taken place.

However, there is an equivalent way of gauge-fixing, which allows to take from the beginning the limit $ \xi \to 0$. This requires the introduction of a Nakanishi-Lautrup \cite{Nakanishi:1966zz, lautrup1967} auxiliary field $ B_{\mu} $. The
alternative gauge-fixing Lagrangian may be written in the form \begin{equation}\label{eq:25}
    \mathcal{L}_{\text{L} } =\sqrt{\bar{g}} B_{\mu} \left (  \bar{\mathsf{D}}_{\nu} h^{\mu \nu} - \frac{1}{2} \bar{\mathsf{D}}^{\mu } h_{\alpha}^{\alpha} \right),
\end{equation}
Integrating the exponential $ \exp ( i \int \mathop{d^{4} x} \mathcal{L}_{\text{L}} )$ over the $ B_{\mu} $ field yields the delta function
\begin{equation}\label{eq:26}
   \delta \left[
   \bar{\mathsf{D}}_{\nu} h^{\mu \nu} - \frac{1}{2} \bar{\mathsf{D}}^{\mu} h_{\alpha}^{\alpha} \right],
\end{equation}
which imposes the Landau background gauge condition.

The relevant Faddeev-Popov ghost Lagrangian at one-loop order \cite{Faddeev:1967fc}  is given by
\begin{equation}\label{eq:27}
    \mathcal{L}_{\text{gh}} = \sqrt{\bar{g}} \eta^{*}{}^{ \mu} \left[ \bar{\mathsf{D}}_{\alpha} \bar{\mathsf{D}}^{\alpha} \tensor{\eta}{_{\mu }} - \bar{R}_{\mu \nu} \eta^{\nu} \right]
   ,
\end{equation}
where  $\bar{R}_{\mu \nu} $ is the background Ricci tensor  and $ \eta^{\mu} $, $ \eta^{*}{}^{\mu} $ are vector ghost  fields.

Proceeding in this way, it turns out that the effective Lagrangian
\begin{equation}\label{eq:28}
    \mathcal{L}_{\text{eff}} (\bar{g}, \bar{\phi} , h, \varphi , B , \eta^{*}{}^{} , \eta ) = \mathcal{L}^{\text{inv} } ( \bar{g}  + \kappa h, \bar{\phi} +  \varphi ) + \mathcal{L}_{\text{L}} + \mathcal{L}_{\text{gh}}
\end{equation}
is invariant under the  BRST transformations
\begin{subequations}\label{eq:209}
    \begin{align}\label{eq:209a}
        & \mathop{\mathsf{s}} \bar{g}_{\mu \nu} = 0, \quad \mathop{\mathsf{s}} \bar{\phi} = 0; \\
        & \mathop{\mathsf{s}} h_{\mu \nu} = \mathsf{D}{_{\mu}} \eta_{\nu} + \mathsf{D}_{\nu} \eta_{\mu}, \quad \mathop{\mathsf{s}} \varphi = \kappa  \partial_{\tau} (\bar{\phi} + \varphi )  \eta^{\tau} ; \\
        & \mathop{\mathsf{s}} \eta^{\mu} = \kappa \eta^{\nu} \mathsf{D}_{\nu} \eta^{\mu}
        ; \quad \mathop{\mathsf{s}} \eta^{*}_{\mu} = - B_{\mu} ; \\
        & \mathop{\mathsf{s}} B_{\mu} =0,
    \end{align}
\end{subequations}

In order to derive the corresponding Slavnov-Taylor identity \cite{taylor:1976} which reflects this symmetry, it is useful to introduce the external sources  $ G^{\star \mu \nu}$  and $ C^{\star \mu} $ coupled respectively to the BRST transformations of the graviton field $h_{\mu \nu} $ and the ghost field $ \eta_{\mu} $ \cite{Kluberg-Stern:1974iel}. These sources have dimensions of mass$^2$, ghost number $-1$ and $-2$ and are invariant under BRST transformations, so that
\begin{equation}\label{eq:210}
    \mathop{\mathsf{s}} G^{\star \mu \nu} = \mathop{\mathsf{s}} C^{\star \mu} =0.
\end{equation}

Using Eq.~\eqref{eq:28}, the full Lagrangian $ \mathcal{L}_{\text{full}} $ becomes 
\begin{equation}\label{eq:211}
    \mathcal{L}_{\text{full}} = \mathcal{L}_{\text{eff}} + G^{\star \mu \nu} \mathop{\mathsf{s}} h_{\mu \nu} + C^{\star \mu} \mathop{\mathsf{s}} \eta_{\mu}.
\end{equation}
From the BRST symmetry of the theory, one thus obtains the ST identity for the action $ \Gamma_{0}(\bar{g}, \bar{\phi} ; h, \varphi , B, \eta^{*}{} , \eta ; G^{\star} , C^{\star} ) = \int \mathop{d^{4} x} \mathcal{L}_{\text{full}}$:
\begin{equation}\label{eq:212}
    \begin{split}
        \int & \mathop{d^{4} x}  {\left [
    \frac{\delta \Gamma_{0}}{\delta h_{\mu \nu}} \mathop{\mathsf{s}} h^{\mu \nu}
    + \frac{\delta \Gamma_{0}}{\delta \eta^{\mu} } \mathop{\mathsf{s}}  \eta^{\mu}
    + \frac{\delta \Gamma_{0}}{\delta \eta^{*}{}^{\mu} } \mathop{\mathsf{s}} \eta^{*}{}^{\mu}
    +
    \frac{\delta \Gamma_{0}}{\delta \varphi } \mathop{\mathsf{s}} \varphi
\right ]}
\\
={}&
    \int \mathop{d^{4} x} \left [
    \frac{\delta \Gamma_{0}}{\delta h_{\mu \nu}}
    \frac{\delta \Gamma_{0}}{\delta G^{\star \mu \nu}}
    + \frac{\delta \Gamma_{0}}{\delta \eta_{\mu} }
    \frac{\delta \Gamma_{0}}{\delta C^{ \star \mu} }
    - B_{\mu} \frac{\delta \Gamma_{0}}{\delta \eta^{*}_{\mu} }
    + \kappa (\partial^{\rho} \phi)\eta_{\rho}
    \frac{\delta \Gamma_{0}}{\delta \varphi }
\right ]=0.
\end{split}
\end{equation}

Moreover, one also gets from Eq.~\eqref{eq:211} the relation
\begin{equation}\label{eq:213}
    \frac{\delta \Gamma_{0}}{\delta \eta^{*}{}^{\mu}} = \sqrt{\bar{g}} {\left [ D^{\nu} \frac{\delta \Gamma_{0}}{\delta G^{\star \mu \nu}} - \frac{1}{2} \bar{g}^{\rho \sigma}\bar{\mathsf{D}}_{\mu} \frac{\delta \Gamma_{0} }{\delta G^{\star \rho \sigma}} \right ]} =0,
\end{equation}
which allows to simplify the ST identity. Substituting Eq.~\eqref{eq:213} into Eq.~\eqref{eq:212} leads to the equation 
\begin{equation}\label{eq:214}
    \int \mathop{d^{4} x} \left [
    \frac{\delta \Gamma^{\prime}_{0}}{\delta h_{\mu \nu}}
    \frac{\delta \Gamma^{\prime}_{0}}{\delta G^{\star \mu \nu}}
    + \frac{\delta \Gamma^{\prime}_{0}}{\delta \eta_{\mu} }
    \frac{\delta \Gamma^{\prime}_{0}}{\delta C^{ \star \mu} }
    + \kappa (\partial^{\rho} \phi)\eta_{\rho}
    \frac{\delta \Gamma^{\prime}_{0}}{\delta \varphi }
\right ]=0.
\end{equation}
where the action $ \Gamma_{0}^{'} $   is obtained from $ \Gamma_{0} $ by subtracting the Landau gauge-fixing term \eqref{eq:25}:
\begin{equation}\label{eq:215}
    \Gamma^{\prime}_{0} = \Gamma_{0} - \int \mathop{d^{4} x} \sqrt{\bar{g}} B^{\mu} \left( \bar{\mathsf{D}}^{\nu} h_{\mu \nu} - \frac{1}{2} \bar{\mathsf{D}}_{\mu} h_{\sigma}^{\sigma} \right).
\end{equation}

It may be shown, by using a similar reasoning to that in \cite{taylor:1976}, that the Eq.~\eqref{eq:214} holds to all orders for the generating functional $ \Gamma^{'} $ of one-particle irreducible Green functions, which reduces at lowest order to $ \Gamma_{0}^{'} $. Thus, one gets  
\begin{equation}\label{eq:214a}
    \int \mathop{d^{4} x} \left [
    \frac{\delta \Gamma^{\prime}}{\delta h_{\mu \nu}}
    \frac{\delta \Gamma^{\prime}}{\delta G^{\star \mu \nu}}
    + \frac{\delta \Gamma^{\prime}}{\delta \eta_{\mu} }
    \frac{\delta \Gamma^{\prime}}{\delta C^{ \star \mu} }
    + \kappa (\partial^{\rho} \phi)\eta_{\rho}
    \frac{\delta \Gamma^{\prime}}{\delta \varphi }
\right ]=0.
\end{equation}
The above ST identity, which has an universal form, is most useful to prove, in a gauge invariant way, the renormalizability of quantum gravity. 
Although quantum gravity is non-renormalizable by power counting, this theory is renormalizable in the modern sense that the divergences can be cancelled by introducing all possible counterterms allowed by its gauge symmetry \cite{Gomis:1995jp, Barvinsky:2017zlx, Lavrov:2019nuz}.

\section{Counterterm Lagrangian in the Landau background gauge}
The gauge invariance of the background effective action implies that the counterterm Lagrangian may be expressed in terms of invariant functions under the background gauge transformations. Altogether, there are nine independent structures which exhaust all possible independent terms expected from dimensional analysis, symmetry and the form of the interaction terms, namely 
\begin{equation}\label{eq:n38}
    \begin{split}
    & \bar{R}^{2} ; \quad \bar{R}_{\mu \nu}^{2} ; \quad  (\bar{\mathsf{D}}_{\mu} \bar{\mathsf{D}}^{\mu} \bar{\phi} )^{2} ; 
    \quad \bar{R} \partial_{\mu} \bar{\phi} \partial^{\mu} \bar{\phi} ; \quad \bar{R}^{\mu \nu} \partial_{\mu} \bar{\phi} \partial_{\nu} \bar{\phi}; 
    \quad 
    ( \partial_{\mu} \bar{\phi} \partial^{\mu} \bar{\phi} )^{2}; 
    \\
    & \bar{R} \bar{\phi} \bar{\mathsf{D}}_{\mu} \bar{\mathsf{D}}^{\mu} \bar{\phi} ; \quad \bar{R}^{\mu \nu} \bar{\phi} \bar{\mathsf{D}}_{\mu} \bar{\mathsf{D}}_{\nu} \bar{\phi} ; \quad \bar{\phi} \partial^{\mu} \bar{\phi} \partial^{\nu} \bar{\phi} \bar{\mathsf{D}}_{\mu} \bar{\mathsf{D}}_{\nu} \bar{\phi}.
    \end{split}
\end{equation}
Here, we have taken into account that, owing to the identity~\cite{tHooft:1974toh, *thooft1993}
\begin{equation}\label{eq:n39}
    \int \mathop{d^{4} x} \sqrt{\bar{g}} ( \bar{R}_{\mu \nu \rho \sigma}^{2} + \bar{R}^{2} - 4 \bar{R}_{\mu \nu}^{2} )=0, 
\end{equation}
the square of the Riemann tensor $ \bar{R}_{\mu \nu \rho \sigma}^{2} $ is not an independent quantity. In our model, characterized by the Landau gauge fixing term \eqref{eq:25}, the last three structures in the Eq.~\eqref{eq:n38} do not actually appear.

The one-loop background counterterm Lagrangian in the Landau gauge may be expressed in terms of  six independent structures  as
\begin{equation}\label{eq:31}
    \begin{split}
    \mathcal{L}_{\text{CT}}
    = \frac{\sqrt{\bar{g}}}{16 \pi^{2} \epsilon} & \big [ c_{1} \bar{R}^{2} + c_{2}  \bar{R}_{\mu \nu} \bar{R}^{\mu \nu} + \kappa^{2} c_{3}  ( \bar{\mathsf{D}}_{\mu} \bar{\mathsf{D}}^{\mu} \bar{\phi} )^{2}  + \kappa^{2} c_{4}  \bar{R} \partial_{\mu} \bar{\phi} \partial^{\mu} \bar{\phi} + \kappa^{2} c_{5}  \bar{R}^{\mu \nu} \partial_{\mu} \bar{\phi} \partial_{\nu} \bar{\phi} + \kappa^{4} c_{6} ( \partial_{\mu} \bar{\phi} \partial^{\mu} \bar{\phi} )^{2}  \big ],
\end{split}
\end{equation}
where $ \epsilon = (4-D)/2$, $D$ is the spacetime dimension and $ c_{i} $ are dimensionless coefficients.

The Feynman rules necessary for the calculation of the counterterm  Lagrangian  are derived in Appendix A. The tree-level propagators are determined by inverting the quadratic part of the quantum fields present in the Lagrangian \eqref{eq:28}, where the gauge fixing term $ \mathcal{L}_{\text{L}} $ is given in Eq.~\eqref{eq:25}. Because this quadratic part is not diagonal, one gets mixed propagators involving the graviton $h_{\mu \nu} $ and the Nakanishi-Lautrup fields $ B_{\mu} $. However, the $ \langle 0|T B_{\mu} B_{\nu}| 0 \rangle $ propagator vanishes in consequence of the fact that $B_{\mu} $ is a non-dynamical Lagrange multiplier field.

\begin{figure}[ht]
    \centering
    \includegraphics[width=0.55\textwidth]{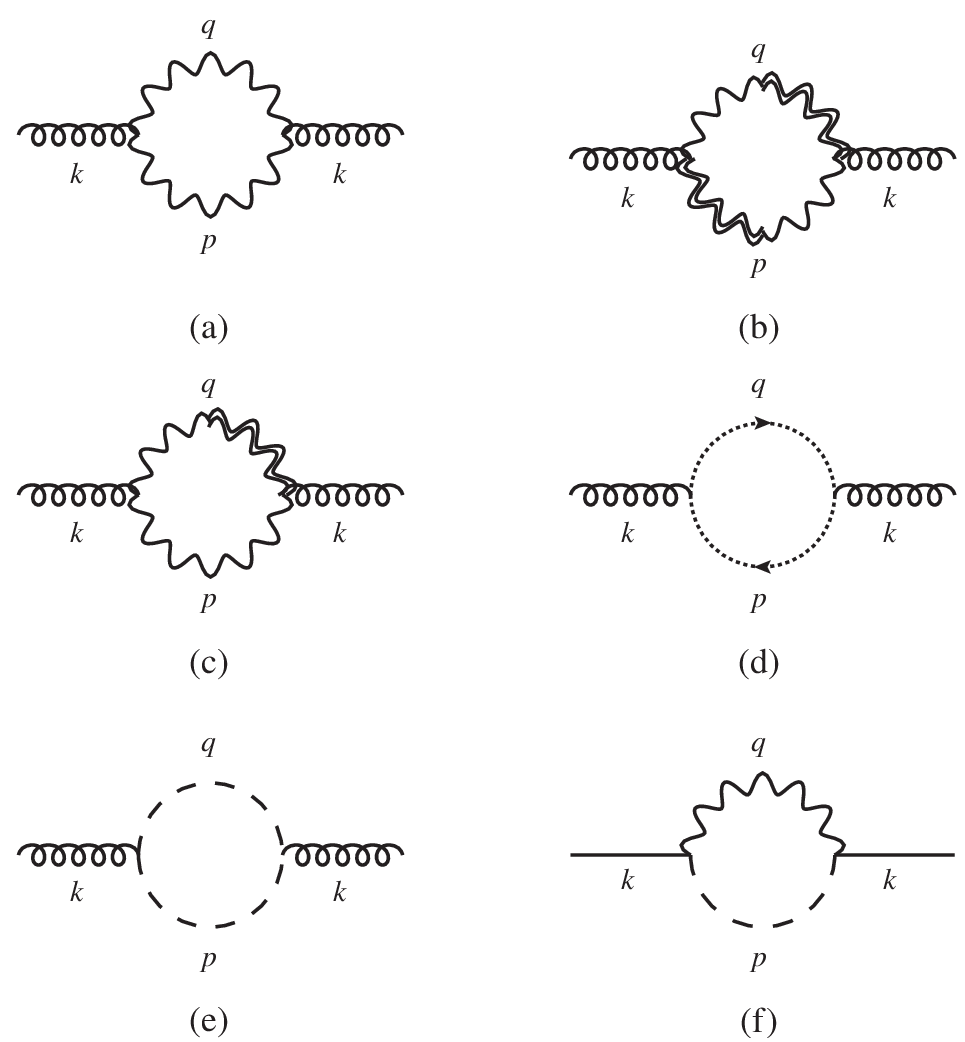}
    \caption{One-loop contributions to $ \left \langle \bar{\mathfrak{g}}^{\mu \nu} \bar{\mathfrak{g}}^{\alpha \beta}\right\rangle $ [(a)-(e)] and to $ \left \langle \bar{\phi} \bar{\phi}\right\rangle $ [(f)] (permutation have been omitted). The wavy, double wavy, curly, dotted, dashed and solid lines denote the quantum graviton $ {h}_{\mu \nu} $, Nakanishi-Lautrup $ B_{\mu} $, background graviton $ \bar{\mathfrak{g}}_{\mu \nu} $, ghost $ \eta^{\mu} $ and $ \eta^{*}{}^{\mu} $, quantum scalar $ \varphi $, and the background scalar $ \bar{\phi} $ fields, respectively. The momenta satisfies the relation: $ q =p+k$.}\label{fig:2point}
\end{figure}

In order to fix the coefficients $c_1$ and $c_2$, we first consider the background graviton self-energy diagrams illustrated in Fig.~\ref{fig:2point}.
As shown in Appendix B, these graphs yield in momentum space  the divergent contribution
\begin{equation}\label{eq:32}
    \bar{\Pi}^{\mu \nu \alpha \beta}(k) = \frac{\kappa^{2} k^{4}}{16 \pi^{2} \epsilon} \ \frac{43}{40}{\left [  T^{\mu \nu} T^{\alpha \beta} + \frac{1}{6} {\left (  T^{\mu \alpha} T^{\nu \beta} + T^{\mu \beta} T^{\nu \alpha}\right )}  \right ]},
\end{equation}
where $ T^{\mu \nu} = k^{\mu} k^{\nu} / k^{2} - \eta^{\mu \nu} $. We note that this expression is transverse with respect to the momentum $k$, which is a consequence of the diffeomorphism invariance of the theory [see Eq.~\eqref{eq:n35}].

We can connect the above result with the counterterm Lagrangian \eqref{eq:31}, by using the relations
\begin{align}\label{eq:33}
    \bar{R} (k) ={}& \kappa k^{2} T^{\mu \nu}  \bar{\mathfrak{g}}_{\mu \nu} + O ( \kappa^{2} \bar{\mathfrak{g}}^{2} ),
    \\\label{eq:34}
    \bar{R}^{\alpha \beta} (k) ={}& \frac{\kappa}{2} \left [ T^{\mu \nu} k^{\alpha} k^{\beta} - \frac{k^{2}}{2} \left ( T^{\alpha \mu} T^{\beta \nu} + T^{\alpha \nu} T^{\beta \mu}\right )\right ] \bar{\mathfrak{g}}_{\mu \nu} + O ( \kappa^{2} \bar{\mathfrak{g}}^{2}  ).
\end{align}
where we have defined a non-flat background field $ \bar{\mathfrak{g}}_{\mu \nu} $ such that
\begin{equation}\label{eq:35}
    \bar{g}_{\mu \nu} = \eta_{\mu \nu } + \kappa \bar{\mathfrak{g}}_{\mu \nu}.
\end{equation}

Then, one may verify that the Eq.\eqref{eq:32} determines the coefficients $ c_{1} $ and $ c_{2} $ as
\begin{equation}\label{eq:36}
    c_{1} = \frac{43}{240} ; \quad c_{2} = \frac{43}{120}.
\end{equation}
Moreover, the divergent part of the scalar self-energy, shown in Fig.~\ref{fig:2point}(f) vanishes in the Landau background gauge, so that
\begin{equation}\label{eq:37}
    c_{3} =0.
\end{equation}

\begin{figure}[ht]
    \centering
    \includegraphics[width=0.8\textwidth]{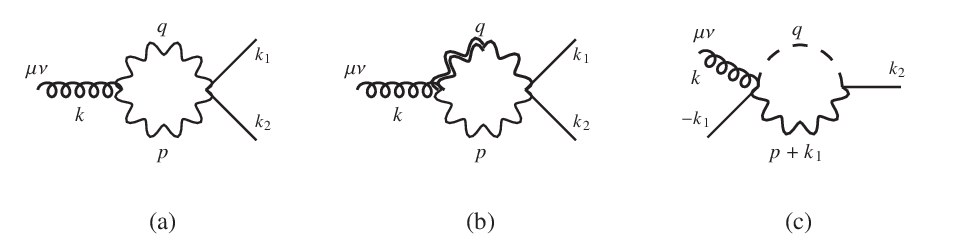}
    \includegraphics[width=0.8\textwidth]{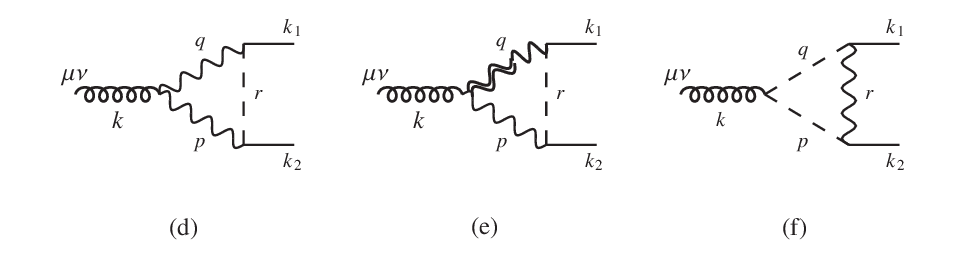}
    \caption{One-loop contributions (permutations have been omitted) to the background $3$-point function $ \langle \bar{\mathfrak{g}}^{\mu \nu} (k) \bar{\phi} (k_1) \bar{\phi} (k_2) \rangle$ [(a)-(f)]. We have that $ q = p+k$,  $ r =p+k-k_1$ and $ k = k_{1} + k_{2} $.}\label{fig:3point}
\end{figure}

In order to fix the coefficients $c_{4} $ and $c_{5} $ in Eq. \eqref{eq:31}, we consider the diagrams illustrated in Fig.~\ref{fig:3point}.
As shown in Appendix B, by using the Eqs. \eqref{eq:33} and \eqref{eq:34}, the divergent contributions of such diagrams determine these  coefficients to be
\begin{equation}\label{eq:38}
    c_{4} = \frac{1}{3} ; \quad c_{5} = - \frac{5}{6} .
\end{equation}

\begin{figure}
    \centering
\includegraphics[width=0.8\textwidth]{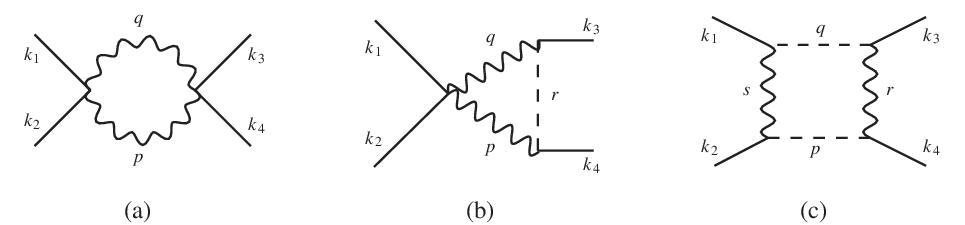} 
\caption{One-loop  contributions (permutations have been omitted) to the background $4$-point function $ \left \langle \bar{\phi} (k_{1} )\bar{\phi} (k_{2} )\bar{\phi} (k_{3} )\bar{\phi} (k_{4} )\right\rangle$ [(a)-(c)]. Here, $ q = p+k_{1} + k_{2} $, $ r = p + k_{4}  $ and $ s = p +k_{2} $.}\label{fig:scalar_scalar}
\end{figure}

Finally, we consider the divergent parts of the diagrams shown in Fig.~\ref{fig:scalar_scalar}, which fix the coefficient $c_6$.
These contributions are evaluated in Appendix B and determine this coefficient as 
\begin{equation}\label{eq:39}
    c_{6} = \frac{1}{4} .
\end{equation}

Substituting the coefficients \eqref{eq:36}--\eqref{eq:39} in the Eq.~\eqref{eq:31}, we obtain at one-loop order that the off-shell background counterterm Lagrangian in the Landau  gauge  is given by 
\begin{equation}\label{eq:310}
     \mathcal{L}_{\text{CT}}
 = \frac{\sqrt{\bar{g}}}{16 \pi^{2} \epsilon} \left[ \frac{43}{240} \left(\bar{R}^{2} + 2\bar{R}_{\mu \nu} \bar{R}^{\mu \nu} \right)
 + \frac{1}{3}\kappa^{2} 
 \bar{R}  \partial_{\mu} \bar{\phi} \partial^{\mu} \bar{\phi} 
- \frac{5}{6} \kappa^{2} 
\bar{R}^{\mu \nu} \partial_{\mu} \bar{\phi} \partial_{\nu} \bar{\phi}  
+ \frac{1}{4}\kappa^{4} ( \partial_{\mu} \bar{\phi} \partial^{\mu} \bar{\phi} )^{2} 
\right].   
\end{equation}
This result coincides with that found in Ref. \cite{Frenkel:2026zwc} in a general background gauge, when all gauge parameters are set, at the end of the computations, equal to zero. We note that the present method of calculation, in the Landau gauge, is much simpler and direct.

In order to obtain the on-shell  background counterterm Lagrangian, we use Einstein’s equation 
\begin{equation}\label{eq:311}
    \bar{R}_{\mu \nu} - \frac{1}{2} \bar{R} \bar{g}_{\mu \nu} = \frac{\kappa^{2}}{2} \mathfrak{T}_{\mu \nu} ,
\end{equation}
where the energy-momentum tensor $ \mathfrak{T}_{\mu \nu} $ corresponding to the Lagrangian~\eqref{eq:21} is given by 
\begin{equation}\label{eq:312}
    \mathfrak{T}_{\mu \nu} = \frac{1}{2} \bar{g}_{\mu \nu} \partial_{\alpha} \bar{\phi} \partial^{\alpha} \bar{\phi} - \partial_{\mu} \bar{\phi} \partial_{\nu} \bar{\phi}.  
\end{equation}
From the above equations, we obtain the on-shell relations 
\begin{equation}\label{eq:313}
    \bar{R}_{\mu \nu} = - \frac{\kappa^{2}}{2} \partial_{\mu} \bar{\phi} \partial_{\nu} \bar{\phi} ; \quad \bar{R} = - \frac{\kappa^{2}}{2} \partial_{\mu} \bar{\phi} \partial^{\mu} \bar{\phi} .
\end{equation}

Using these relations in Eq.\eqref{eq:310}, the on-shell counterterm Lagrangian turns out to be 
\begin{equation}\label{eq:314}
    \mathcal{L}_{\text{CT}} \big |_{\text{on-shell}} = \frac{\sqrt{\bar{g}} \kappa^{4}}{16 \pi^{2} \epsilon} \frac{203}{320} ( \partial_{\mu} \bar{\phi} \partial^{\mu} \bar{\phi} )^{2}. 
\end{equation}
This result is consistent with that obtained on-shell in the Feynman background gauge in Ref.~\cite{tHooft:1974toh, *thooft1993}.

\section{Background effective action}

The action corresponding to the Lagrangian $ \mathcal{L}^{\text{inv} }  $ [see Eq.~\eqref{eq:21}] is also invariant under the background gauge transformations 
    \begin{align}\label{eq:n31}
        \delta \bar{g}_{\mu \nu} = \kappa( \bar{\mathsf{D}}_{\mu} \epsilon_{\nu} + \bar{\mathsf{D}}_{\nu} \epsilon_{\mu} ); \quad \delta \bar{\phi} = \kappa \epsilon^{\alpha} \partial_{\alpha} \bar{\phi} 
        \intertext{and}
        \delta h_{\mu \nu} = \kappa( \epsilon^{\gamma} \partial_{\gamma} h_{\mu \nu} + h_{\mu \gamma} \partial_{\nu} \epsilon^{\gamma} + h_{\nu \gamma} \partial_{\mu} \epsilon^{\gamma} ); \quad \delta \varphi = \kappa \epsilon^{\alpha} \partial_{\alpha} \varphi,
    \end{align}
    where $ \epsilon^{\alpha} $ is an arbitrary infinitesimal parameter. The actions corresponding to $ \mathcal{L}_{\text{L}} $ and $ \mathcal{L}_{\text{gh}} $ [see Eqs.~\eqref{eq:25} and \eqref{eq:27}] are invariant under these transformations, provided the fields $ B_{\mu} $, $ \eta^{*}_{\mu} $ and $ \eta_{\mu} $ transform as a covariant vector field 
\begin{equation}\label{eq:n32}
        \delta V_{\mu} = - \epsilon^{\sigma} \partial_{\sigma} V_{\mu} - V_{\sigma } \partial_{\mu} \epsilon^{\sigma} .
\end{equation}

The background effective action $ \bar{\Gamma} $ is obtained by evaluating the effective action $ \Gamma $ at vanishing quantum fields and sources, namely 
\begin{equation}\label{eq:n33}
    \bar{\Gamma} (\bar{g}, \bar{\phi} ) = \Gamma (\bar{g}, \bar{\phi} ; 0;0).
\end{equation}
The gauge invariance of this action under the background gauge transformations may be expressed in coordinate space as 
\begin{equation}\label{eq:n34}
    \int \mathop{d^{4} x} \left(\frac{\delta \bar{\Gamma}}{\delta \bar{g}_{\mu \nu}} \delta \bar{g}_{\mu \nu}(x) + \frac{\delta \bar{\Gamma}}{\delta \bar{\phi}} \delta \bar{\phi}(x) \right)=0.
\end{equation}

Taking the functional derivative of this equation with respect to $\bar{\mathfrak{g}}_{\alpha\beta}$, evaluated at $\bar{\mathfrak{g}}=0$ and $\bar{\phi}=0$, one obtains, after a Fourier transformation to momentum space, the relation
\begin{equation}\label{eq:n35}
k^{\mu}\bar{\Pi}_{\mu\nu,\alpha\beta}(k)=0,
\end{equation}
where we have used the relation $\bar{g}_{\mu\nu} = \eta_{\mu\nu} + \kappa\bar{\mathfrak{g}}_{\mu\nu}$ together with Eq.~\eqref{eq:n31}. This equation shows that the background graviton self-energy is a transverse quantity.

Another Ward identity may be obtained by taking functional derivative of the Eq.~\eqref{eq:n34} with respect to $ \bar{\mathfrak{g}}_{\alpha \beta} (y) $ and $ \bar{\mathfrak{g}}_{\rho \sigma} (z)$, evaluated at $ \bar{\mathfrak{g}} =0$ and $ \bar{\phi} =0$. We then obtain in momentum space the expression 
\begin{equation}\label{eq:nn35}
    k^{\mu} \bar{V}_{\bar{\mathfrak{g}}_{\mu \nu} \bar{\mathfrak{g}}_{\alpha \beta} \bar{\mathfrak{g}}_{\rho \sigma}} (k,p,q) = \frac{\kappa}{2} { \left\{ p^{\lambda} {\left [ \eta_{\nu \rho} \bar{\Pi}_{\lambda \sigma \alpha \beta }(q) + \eta_{\nu \sigma} \bar{\Pi}_{\lambda \rho \alpha \beta } (q) \right ]} 
    -p_{\nu} \bar{\Pi}_{\alpha \beta \rho \sigma} (q) + (p, \alpha \beta ) \leftrightarrow (q , \rho \sigma ) \right\}}, 
\end{equation}
which relates the four-divergence of the background graviton $3$-point vertex to the background graviton self-energies. 

Moreover, taking the functional derivative of the Eq.~\eqref{eq:n34} with respect to $ \bar{\phi} (y)$ and $ \bar{\phi} (z)$ evaluated at $ \bar{\mathfrak{g}} = \bar{\phi} =0$,  we obtain in momentum space the relation 
\begin{equation}\label{eq:n36}
    k^{\mu}  \bar{V}_{\bar{\mathfrak{g}}_{\mu \nu} \bar{\phi} \bar{\phi}} (k,p,q)=  \frac{\kappa}{ 2} {\left [ q_{\nu} \bar{\Sigma} (p) + p_{\nu}  \bar{\Sigma}(q) \right ]} ,
\end{equation}
where $ \bar{\Sigma}$ is the background scalar self-energy function. In the previous section, we have verified, at one-loop order, that the divergent part of this function vanishes  in the Landau gauge. This implies that the background graviton-scalar-scalar vertex satisfies the transversality condition 
\begin{equation}\label{eq:n37}
    k^{\mu}  \bar{V}^{\text{div}}_{\bar{\mathfrak{g}}_{\mu \nu} \bar{\phi} \bar{\phi}} (k,p,q)=0.
\end{equation}
Conversely, this equation implies, by using Eq.~\eqref{eq:n36}, that the scalar self-energy in the Landau background gauge must be finite at this order.  
The transversality property may be explicitly verified at one loop order  by noting that the  divergent part of the above vertex can be expressed, as we have seen in the previous section, in terms of  the gauge-invariant structures $ \bar{R} \partial_{\alpha} \bar{\phi} \partial^{\alpha} \bar{\phi} $ and $ \bar{R}^{\alpha \beta } \partial_{\alpha} \bar{\phi} \partial_{\beta} \bar{\phi} $. Taking functional derivatives of these structures with respect to $ \bar{\mathfrak{g}}_{\mu \nu}, \bar{\phi}, \bar{\phi} $ evaluated at $ \bar{\mathfrak{g}} = \bar{\phi} = 0$, one obtains transverse results due to the form of the Ricci scalar and the Ricci tensor shown in Eqs.~\eqref{eq:33} and \eqref{eq:34}.

A similar feature might occur at higher orders. Drawing on the behaviour at one-loop order, one may suppose that at two loop order the divergent part of the background graviton-scalar-scalar vertex could also be expressed in terms of gauge-invariant structures like 
\begin{equation}\label{eq:n410}
    \mathcal{L}_{\text{CT}}^{(2)} = \kappa^{4} {\left [ d_{1} ( D_{\alpha} D^{\alpha} R) \partial_{\beta} \bar{\phi} \partial^{\beta} \bar{\phi} + d_{2} ( D_{\alpha }D_{\beta} \bar{R}^{\alpha \beta } ) \partial_{\rho} \bar{\phi} \partial^{\rho} \bar{\phi} + \cdots\right ]}.    
\end{equation}
By taking a functional derivative of such structures with respect to $ \bar{\mathfrak{g}}_{\mu \nu} $, evaluated at $ \bar{\mathfrak{g}} =0$, one gets a vanishing result unless this derivative acts directly on the background field appearing in the scalar Ricci $R$ and in the Ricci tensor $ \bar{R}^{\alpha \beta} $. Then, as explained above, these contributions will be transverse.                               

\begin{figure}[ht]
    \includegraphics[width=0.85\textwidth]{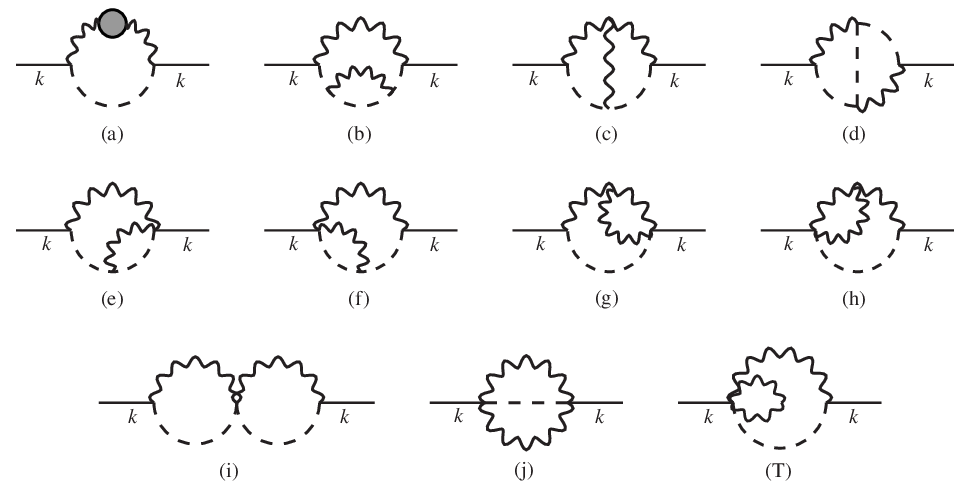}
    \caption{Two-loop contributions to the background scalar self-energy [(a)-(j)]. Diagram (T) represents tadpoles contributions, which vanish in dimensional regularization. The blob in Fig. 4(a) denotes the one-loop quantum graviton self-energy, which is similar to that shown in Fig.~\ref{fig:2point}.}\label{fig:2loop} 
\end{figure}

Thus, such a reasoning suggests that in the Landau background gauge, due to the transversality condition \eqref{eq:n37}, the Ward identity \eqref{eq:n36} may imply the finiteness of the massless  scalar self-energy function also at higher orders.
The Feynman diagrams contributing at two-loop order to this function are shown in Fig.~\ref{fig:2loop}.
The evaluation of the divergent parts of these diagrams, which is rather involved, will be subsequently addressed.

\section{Conclusion}
The Landau gauge is very useful as it leads to a significant simplification of calculations as well as to an improved understanding of various phenomena in particle  physics.
It has been much used in QED to study quantum electromagnetic processes \cite{itzyksonQuantumFieldTheory1980} and in QCD in the analysis  of the properties of quarks and gluons \cite{Berges:2000ew, Binosi:2009qm, Maas:2011se}. In quantum gravity, much work has also been done in the Landau gauge in the context of the functional renormalization (see \cite{Kawasaki:1981zw, Eichhorn:2010tb, Gies:2015tca, Ohta:2016jvw, Ohta:2016npm, Christiansen:2017bsy} and references therein). 

In this work, we employed the background field approach, which has the important virtue of preserving the diffeomorphism invariance  of the theory at all stages of the calculations.
We examined some basic aspects of the BRST symmetry  of quantum gravity in the Landau background gauge and analyzed the Ward identities which reflect the invariance of the background effective action. We have calculated, at one-loop order, the gauge invariant off-shell effective action of this theory [Eq.\eqref{eq:310}] and shown that, on-shell, this action yields a result [Eq.\eqref{eq:314}]
which is in agreement with the one obtained by  ‘t  Hooft and Veltman in the  Feynman background gauge \cite{tHooft:1974toh, *thooft1993}. This result is expected in view of the Kallosh-DeWitt theorem, which asserts that the on-shell effective action should be independent of the gauge-fixing parameters \cite{Kallosh:1974yh, DeWitt:1967ub}.

An interesting outcome of the present  analysis in quantum gravity concerns the possible finiteness, in the Landau background gauge, of the off-shell scalar self-energy at higher orders [On symmetry and power counting grounds, it is evident that the on-shell ($ k^{2} =0$) massless scalar self-energy vanishes at all orders]. We indicated  that such a property,  which was explicitly verified at one-loop order, may  follow at higher orders  in consequence of the background Ward identity involving the background graviton-scalar-scalar vertex [see Eq.~\eqref{eq:n36}]. This inference could be attested at two loop order by explicitly computing the contributions coming from the scalar self-energy diagrams shown in Fig.~\ref{fig:2loop}. We are currently  considering this issue.

\begin{acknowledgments}
We thank Fernando Brandt for helpful comments. 
S.\ M.-F.\ thanks Gastão Krein for useful conversations. J.\ F.\ thanks CNPq (Brazil) for financial support.  S.\ M.-F.\ thanks FAPESP for financial support. This study was financed, in part, by the São Paulo Research Foundation (FAPESP), Brasil, Process No. 2025/16156-7.
\end{acknowledgments}

\appendix

\section{Feynman rules} \label{sec:FR}
In this appendix, we show the Feynman rules required to compute the one-loop counterterms in the Einstein-Hilbert theory coupled to a scalar field in the background Landau gauge. Some Feynman rules are derived in the Ref.~\cite{Frenkel:2026zwc} for a general background gauge. 

Here, we follow the usual definitions and conventions used in Ref. \cite{tHooft:1974toh, *thooft1993}:
\begin{align}\label{eq:A0}
        \bar{R}^{\mu}{}_{\nu \alpha \beta} ={}& -\partial_{\beta} \bar{\Gamma}^{\mu}_{\nu \alpha}
+ \partial_{\alpha} \bar{\Gamma}^{\mu}_{\nu \beta}
+ \bar{\Gamma}^{\mu}_{\gamma \alpha} \bar{\Gamma}^{\gamma}_{\nu \beta}
- \bar{\Gamma}^{\mu}_{\gamma \beta} \bar{\Gamma}^{\gamma}_{\nu \alpha} \\
        \bar{R}_{\mu \nu \alpha \beta} ={}& \bar{g}_{\mu \gamma} \bar{R}^{\gamma}{}_{\nu \alpha \beta},
\quad
\bar{R}_{\nu \beta} = \bar{R}^{\mu}{}_{\nu \beta \mu},
\quad
\bar{R}  = \bar{R}_{\nu \beta} \bar{g}^{\nu \beta}.
\end{align}
Moreover, we will use that
\begin{equation}\label{eq:A1}
    \bar{g}_{\mu \nu} = \eta_{\mu \nu} + \kappa \bar{\mathfrak{g}}_{\mu \nu},
\end{equation}
where $ \eta_{\mu \nu} $ is a flat background and $ \bar{\mathfrak{g}}_{\mu \nu} $ is the background graviton field.

In order to derive the Feynman rules, we used the definitions \eqref{eq:22} and \eqref{eq:A1}\allowdisplaybreaks
\begin{subequations}\label{eq:A2}
    \begin{align}\label{eq:A2a}
        g^{\mu \nu} ={}& \eta^{\mu \nu} - \kappa \bar{\mathfrak{g}}^{\mu \nu}
        - \kappa \eta^{\mu \alpha} \eta^{\nu \beta} h_{\alpha \beta}
        + \kappa^{2} \eta^{\mu \alpha} \bar{\mathfrak{g}}^{\nu \beta} h_{\alpha \beta}
        + \kappa^{2} \bar{\mathfrak{g}}^{\mu \alpha} \eta^{\nu \beta} h_{\alpha \beta}
        + \kappa^{2} h^{\mu}_{\alpha} h^{\alpha \nu}
        , \\
        \sqrt{g} ={}&
        1 + \frac{\kappa}{2}  \bar{\mathfrak{g}}_{\mu}^{\mu}
        + \frac{\kappa }{2} \eta^{\mu \nu} h_{\mu \nu}
        - \frac{\kappa^{2} }{2} \bar{\mathfrak{g}}^{\mu \nu} h_{\mu \nu}          + \frac{ \kappa^{2} }{4}
        \bar{\mathfrak{g}}_{\mu}^{\mu} \eta^{\alpha \beta} h_{\alpha \beta}
        + \frac{\kappa^{2}}{8} \left ( h_{\alpha}^{\alpha}{}^{2} -2h^{\alpha}_{\beta} h^{\beta}_{\alpha} \right ), \\
        \bar{\Gamma}{_{\mu \nu}^{\lambda}} ={}&
        \frac{\kappa}{2} \eta^{\lambda \rho} ( \partial_{\nu} \bar{\mathfrak{g}}_{\rho \mu} + \partial_{\mu} \bar{\mathfrak{g}}_{\rho \nu} - \partial_{\rho} \bar{\mathfrak{g}}_{\mu \nu} )
        ,\\
        \bar{\mathsf{D}}_{\sigma} h_{\mu \nu} ={}&
 \partial_\sigma h_{\mu\nu}
-
        \frac{\kappa}{2} \eta^{\alpha \rho}
        \left[ ( \partial_{\sigma} \bar{\mathfrak{g}}_{\rho \mu} + \partial_{\mu} \bar{\mathfrak{g}}_{\rho \sigma} - \partial_{\rho} \bar{\mathfrak{g}}_{\mu \sigma} )
            h_{\alpha\nu}
+
         ( \partial_{\sigma} \bar{\mathfrak{g}}_{\rho \nu} + \partial_{\nu} \bar{\mathfrak{g}}_{\rho \sigma} - \partial_{\rho} \bar{\mathfrak{g}}_{\nu \sigma} )
        h_{\mu\alpha}\right],
    \end{align}
\end{subequations}
in the effective EH Lagrangian \eqref{eq:28}.

Consider the perturbation $\bar{g}_{\mu\nu} \rightarrow \tilde{\bar{g}}_{\mu \nu} =\eta_{\mu\nu} + \kappa \tilde{\bar{\mathfrak{g}}}_{\mu\nu}$. One can show that, in momentum space,
\begin{subequations}
\begin{equation}\label{vR}
    \sqrt{\tilde{ \bar{g} }}\tilde{ \bar{R} } 
= \kappa q^2 T^{\alpha \beta } \, \tilde{\bar{\mathfrak{g}}}_{\alpha\beta} + {\cal O}(\kappa^2),
\end{equation}
\begin{equation}\label{vRmn}
    \sqrt{\tilde{ \bar{g} }} \,\tilde{ \bar{R} }^{\mu\nu}
=
\frac{\kappa}{2} \left[
T^{\rho\sigma}(q)\,{{q}^{\mu }{q}^{\nu }}
-\frac{q^2}{2} [T^{\mu \sigma}(q) T^{\rho\nu}(q) + T^{\mu \rho}(q) T^{\sigma\nu}(q) ]
\right] \tilde{\bar{\mathfrak{g}}}_{\rho\sigma}(q) + {\cal O}(\kappa^2).
\end{equation}
\end{subequations}
where
\begin{equation}\label{transvL}
T^{\mu\nu}(q) = \frac{q^\mu q^\nu}{q^2} -  \eta^{\mu\nu}.
\end{equation}
We also have that:
\begin{subequations}\label{Rinvariants}
\begin{equation}\label{RinvA}
\sqrt{\tilde{ \bar g}}\tilde{ R}^2
= \kappa^2 q^4 T^{\alpha \beta }(q) T^{\mu \nu }(q)
\tilde{\bar{\mathfrak{g}}}_{\mu\nu} \tilde{\bar{\mathfrak{g}}}_{\alpha\beta}
+ {\cal O}(\kappa^3),
\end{equation}
\begin{equation}\label{RinvB}
\sqrt{\tilde{ \bar g}}\tilde{ R}_{\mu\nu}\tilde{ R}^{\mu\nu}
= \kappa^2 q^4\biggl[
\frac{1}{4} T^{\alpha \beta }(q) T^{\mu \nu}(q)
+\frac{1}{8} T^{\nu \alpha }(q) T^{\mu \beta}(q)
+\frac{1}{8} T^{\mu \alpha }(q) T^{\nu \beta }(q)
\biggr]
\tilde{\bar{\mathfrak{g}}}_{\mu\nu} \tilde{\bar{\mathfrak{g}}}_{\alpha\beta} + {\cal O}(\kappa^3).
\end{equation}
\end{subequations}

\subsection{Propagators }

The quadratic form in the quantum fields for the graviton sector is given by 
\begin{equation}
    \frac{1}{2}
    \int  
\mathop{d^{4} x}
    \begin{pmatrix}
        h_{\mu\nu} & B_{\alpha}      
    \end{pmatrix}
    \bm{M}  \begin{pmatrix}
        h_{\rho \sigma} \\ B_{\beta}
    \end{pmatrix}
\end{equation}
where the matrix $ \bm{M} $ is equal to 
\begin{equation}\label{eq:defM}
    \bm{M} = \begin{pmatrix}
        A_{\mu \nu \rho \sigma } & B_{\mu \nu \beta} \\
        C_{\alpha \rho \sigma} & D_{\alpha \beta}
    \end{pmatrix},
\end{equation}
where 
\begin{align}\label{eq:quadratic}
    A^{\mu\nu\rho\sigma} ={}
    - \bigg[& 
\partial^{\sigma } \partial^{\nu } \eta^{\rho \mu
   }-\partial^{\rho } \partial^{\sigma } \eta^{\mu \nu }-\frac{1}{2}
   \partial^2 \eta^{\rho \mu } \eta^{\sigma \nu }+\frac{1}{2} \partial^2 \eta^{\rho \sigma }
   \eta^{\mu \nu }\bigg]
\nonumber \\
&+ \mbox{symmetrizations } \mu\leftrightarrow\nu \mbox{ and } \rho\leftrightarrow\sigma
\nonumber \\
&+ \mbox{permutation } (\mu,\nu) \leftrightarrow (\rho,\sigma);
\end{align}
$B_{\mu \nu \beta} = \dfrac{I}{2} ( p_{\beta} \eta_{\mu \nu} - p_{\mu} \eta_{\nu \beta} - p_{\nu} \eta_{\mu \beta} )$, $ C_{\alpha \rho \sigma} = - B_{\rho \sigma \alpha} $ and $ D_{\alpha \beta} = \xi \eta_{\alpha \beta} $.
Note that, in order to obtain the propagators, we have to introduce the parameter $ \xi $ as a regulator. In the end, we can set $ \xi \to 0$ to obtain the propagators in the Landau gauge.

The propagators are given by $i \bm{M}^{-1}$. In order to invert the matrix $ \bm{M} $, we use that 
\begin{equation}\label{eq:formM}
    \bm{M}^{-1} = \begin{pmatrix}
        \bm{X}^{-1} & - \bm{X}^{-1} \bm{B} \bm{D}^{-1} \\
        - \bm{D}^{-1} \bm{C} \bm{X}^{-1} & \bm{D}^{-1} + \bm{D}^{-1} \bm{C} \bm{X}^{-1} \bm{B} \bm{D}^{-1}
    \end{pmatrix}
    ,
\end{equation}
where $ \bm{X} = \bm{A} - \bm{B} \bm{D}^{-1} \bm{C} $ is the Schur complement. 
Then, we need to find $ \bm{D}^{-1}$, $ \bm{X} $ and its inverse. 
The inverse of $ \bm{D} $ is straightforward to obtain: $ D^{-1}_{\alpha \beta} = \eta_{\alpha \beta} / \xi $. Using this, we get
\begin{align}\label{eq:quadraticX}
    X^{\mu\nu\,\alpha\beta} ={}
    - \bigg[& \frac{1}{\xi}\left({\frac{1}{2} \partial^{\mu } \partial^{\nu } \eta^{\alpha
   \beta }-\partial^{\beta } \partial^{\nu } \eta^{\alpha \mu
   }+\frac{1}{2} \partial^{\alpha } \partial^{\beta } \eta^{\mu \nu
   }-\frac{1}{4} \partial^2 \eta^{\alpha \beta } \eta^{\mu \nu }}\right)
\\ & +\partial^{\beta } \partial^{\nu } \eta^{\alpha \mu
   }-\partial^{\alpha } \partial^{\beta } \eta^{\mu \nu }-\frac{1}{2}
   \partial^2 \eta^{\alpha \mu } \eta^{\beta \nu }+\frac{1}{2} \partial^2 \eta^{\alpha \beta }
   \eta^{\mu \nu }\bigg]
\nonumber \\
&+ \mbox{symmetrizations } \mu\leftrightarrow\nu \mbox{ and } \alpha\leftrightarrow\beta
\nonumber \\
&+ \mbox{permutation } (\mu,\nu) \leftrightarrow (\alpha,\beta).
\end{align}

Going to momentum space ($\partial\rightarrow i p$) and computing $i[X^{\mu\nu\alpha\beta}(p)]^{-1}$ with computer algebra (throughout this work we have used Package-{\sf X} \cite{Patel:2015tea}), we obtain
the graviton propagator:
\begin{align}\label{eq:propquantumh}
    {\cal D}_{\mu\nu\,\alpha\beta}^{hh} (p) ={}-\frac{i}{p^2 + i \epsilon }  \Biggl[
   \eta_{\alpha \nu } \eta_{\beta \mu }
+  \eta_{\alpha \mu } \eta_{\beta \nu }
- \eta_{\alpha \beta } \eta_{\mu \nu}
+&\frac{\xi -1}{ { p^2}} \biggl(
p_{\beta } p_{\nu } \eta_{\alpha
    \mu }
+ p_{\beta } p_{\mu } \eta_{\alpha
    \nu }
+ p_{\alpha } p_{\nu } \eta_{\beta
    \mu }
+ p_{\alpha } p_{\mu } \eta_{\beta
    \nu }\biggr)\Biggl].
\end{align}
We can also compute the mixed propagators, which reads 
\begin{equation}\label{eq:mixed}
    \mathcal{D}^{Bh}_{\alpha \, \mu \nu } (p)= \frac{p_{\mu} \eta_{\nu \alpha} + p_{\nu} \eta_{\mu \alpha}}{p^{2} } \quad \text{and} \quad \mathcal{D}^{hB}_{\mu \nu \, \alpha}(p) =  \mathcal{D}^{Bh}_{\alpha \, \mu \nu} (-p).
\end{equation}
Interestingly, the Nakanishi-Lautrup field do not propagates, that is, the $ \langle 0|T B_{\alpha} B_{\beta}| 0 \rangle $ propagator  vanishes: $ \mathcal{D}^{BB}_{\alpha \beta} =0$. 

At this point we may set $\xi = 0$ directly, obtaining the propagators in the Landau gauge. Only the graviton propagator is modified. In this gauge it reads
\begin{align}\label{eqpropquantumh}
    \mathring{{\cal D}}_{\mu\nu\,\alpha\beta}^{hh} (p) ={}-\frac{i}{p^2 + i \epsilon }  \Biggl[
   \eta_{\alpha \nu } \eta_{\beta \mu }
+  \eta_{\alpha \mu } \eta_{\beta \nu }
- \eta_{\alpha \beta } \eta_{\mu \nu}
-&\frac{1}{ { p^2}} \biggl(
p_{\beta } p_{\nu } \eta_{\alpha
    \mu }
+ p_{\beta } p_{\mu } \eta_{\alpha
    \nu }
+ p_{\alpha } p_{\nu } \eta_{\beta
    \mu }
+ p_{\alpha } p_{\mu } \eta_{\beta
    \nu }\biggr)\Biggl].
\end{align}
Similarly, from Eq. \eqref{eq:27} we also can obtain the ghost propagator, which reads
\begin{equation}
{\cal D}_{\mu\nu}(p) = -i \frac{\eta^{\mu\nu}}{p^2+i\epsilon}.
\end{equation}

\subsection{Vertex $\bar{\mathfrak g} h h$ in the background Landau gauge}

The interactions terms $ \bar{\mathfrak{g}} h h $ arises from, where the terms with $ \xi $ (from the gauge fixing) does not appear,
\begin{equation}\label{eq:A14A}
    \mathcal{L}^{(2)}  = -\frac{1}{2} T^{\alpha \mu \nu \beta \gamma \delta} \sqrt{\bar{g}} \bar{\mathsf{D}}_{\alpha} {h}_{\mu \nu} \bar{\mathsf{D}}_{\beta} {h}_{\gamma \delta} - \frac{\sqrt{\bar{g}}}{2} \left [ \bar{R} \left ( \frac{1}{4} h^{2} - \frac{1}{2} h_{\mu \nu} h^{\mu \nu}\right ) + \bar{R}^{\mu \nu} \left ( 2 h^{\alpha}_{\mu} h_{\nu \alpha} - h h_{\mu \nu}\right )\right ] ,
\end{equation}
where $ h \equiv  h_{\mu \nu} \mathfrak{h}^{\mu \nu} $ and
\begin{align}
    T^{\alpha\mu\nu\,\beta\gamma\delta} ={}&
     {H^{(0)}}^{\alpha\mu\nu\beta\gamma\delta}
+
 \frac{1}{2} {H^{(1)}}^{\alpha\mu\nu\beta\gamma\delta}
-\frac{1}{2} {H^{(1)}}^{\alpha\mu\gamma\beta\nu\delta}
+{H^{(1)}}^{\alpha\mu\beta\gamma\nu\delta}
-{H^{(1)}}^{\alpha\mu\nu\delta\gamma\beta}.
\end{align}
with
\begin{equation}\label{eq:H0}
{H^{(0)}}^{\alpha\mu\nu\beta\gamma\delta}\equiv
\frac{1}{2} \eta^{\alpha\beta}\eta^{\mu\gamma} \eta^{\nu\delta}
-\frac{1}{2} \eta^{\alpha\beta}\eta^{\mu\nu} \eta^{\gamma\delta}
+\eta^{\alpha\gamma}\eta^{\mu\nu} \eta^{\beta\delta}
-\eta^{\alpha\delta}\eta^{\mu\gamma} \eta^{\nu\beta}
\end{equation}
and
\begin{equation}
    \begin{split}
        {H^{(1)}}^{\alpha\mu\nu\beta\gamma\delta} ={}&
\kappa\left(-\delta^{\alpha}_{\mu_1}\delta^{\beta}_{\nu_1}  \eta^{\mu\gamma}  \eta^{\nu\delta}
-\eta^{\alpha\beta}  \delta^{\mu}_{\mu_1}\delta^{\gamma}_{\nu_1}  \eta^{\nu\delta}
-\eta^{\alpha\beta}  \eta^{\mu\gamma}  \delta^{\nu}_{\mu_1}\delta^{\delta}_{\nu_1}\right)\mathfrak {\bar{g}}^{\mu_1\nu_1}.
\\ \equiv &
\kappa\; G^{\mu_1\nu_1\,\alpha\mu\nu\beta\gamma\delta}
\mathfrak{\bar{g}}_{\mu_1\nu_1}.
\end{split}
\end{equation}

The total result for the $\bar{\mathfrak g}_{\mu_1\nu_1} h_{\mu_2\nu_2} h_{\mu_3\nu_3}$ interactions vertex is given by
\begin{equation}\label{eq:vertexhhh}
    \vcenter{\hbox{\includegraphics[scale=0.6]{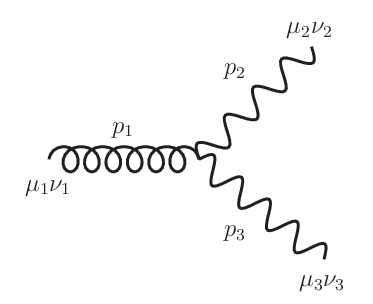}}}
    \quad
    \begin{aligned}     -i\frac{\kappa}{2}
& \Bigg\{
  \delta^{\mu_2}_{\mu} \delta^{\nu_2}_{\nu} \delta^{\mu_3}_{\gamma} \delta^{\nu_3}_{\delta} {p_2}{}_\alpha {p_3}{}_\beta  \bigg[
     \frac{1}{2} \eta^{\mu_1\nu_1} {H^{(0)}}^{\alpha\mu\nu\beta\gamma\delta}
   +    \frac{1}{2} G^{\mu_1\nu_1\,\alpha\mu\nu\beta\gamma\delta}
\\ & -\frac{1}{2} G^{\mu_1\nu_1\,\alpha\mu\gamma\beta\nu\delta}
+G^{\mu_1\nu_1\,\alpha\mu\beta\gamma\nu\delta}
-G^{\mu_1\nu_1\,\alpha\mu\nu\delta\gamma\beta}
\bigg]
\nonumber \\
&
+{H^{(0)}}^{\alpha\mu\nu\beta\gamma\delta}
\Biggr[\Biggl(\frac{1}{2}
  ( \delta_{\mu}^{ \mu_3} \delta_{\nu}^{\nu_3} {{p}_1}{}_{\gamma }
   {p_3}{}_{\alpha } \delta_{\beta}^{ \nu_1} \delta_{\delta}^{ \mu_2} \eta^{\mu_1\nu_2}
   + \delta_{\mu}^{\mu_3} \eta^{\nu_1\nu_2} \delta_{\nu}^{\nu_3} {{p}_1}{}_{\beta } {p_3}{}_{\alpha }
   \delta_{\gamma}^{ \mu_1} \delta_{\delta}^{ \mu_2}
\\ & - \delta_{\mu}^{\mu_3} \delta_{\nu}^{\nu_3}
   {{p}_1}^{\nu_2} {p_3}{}_{\alpha } \delta_{\beta}^{ \mu_1} \delta_{\gamma}^{ \mu_2} \delta_{\delta}^{ \nu_1}
   )
   +
    \gamma \leftrightarrow \delta\Biggr)
+ (\alpha,\mu,\nu) \leftrightarrow(\beta,\gamma,\delta)
\Biggl]
   \\
   & + {p}_1^2 \, T^{\mu_1\nu_1}({p}_1)
\left(
 \frac{1}{4} \eta^{\mu_2\nu_2} \eta^{\mu_3\nu_3} - \frac{1}{2}\eta^{\mu_2\mu_3}\eta^{\nu_2\nu_3}\right)
 \nonumber \\ &+
 \frac{1}{2}\left[
T^{\mu_1\nu_1}({p}_1) {{p}_1}^\mu {{p}_1}^\nu - \frac{{p}_1^2}{2}\left(
T^{\mu\mu_1}({p}_1) T^{\nu\nu_1}({p}_1) + T^{\nu\mu_1}({p}_1) T^{\mu\nu_1}({p}_1)
\right)
\right]
\nonumber\\ &
 \left(
2 \eta^{\nu_3\mu_2} \delta_{\mu}^{\nu_2} \delta_{\nu}^{\mu_3}-\eta^{\mu_2\nu_2}\delta_\mu^{\mu_3}\delta_\nu^{\nu_3}
\right)
\bigg\}
+ \mbox{ symmetrizations } \mu_i\leftrightarrow\nu_i
\nonumber \\
&+ \mbox{ permutation } (\mu_2,\nu_2,p_2) \leftrightarrow (\mu_3,\nu_3,p_3).
\end{aligned}
\end{equation}

\subsection{Nakanishi-Lautrup vertex $B_{\mu} \bar{\mathfrak{g}}_{\alpha \beta } h_{\sigma \rho } $}
The Nakanishi-Lautrup field $ B_{\mu} $ in Eq.~\eqref{eq:25} leads to the following interactions terms, which are relevant to the one-loop order: 
\begin{equation}\label{eq:}
    \begin{split}
 & \kappa B_{\mu}  
\left [
    \frac{1}{2} \bar{\mathfrak{g}}^{\alpha \beta}  \left[ \eta^{\mu \alpha} \eta^{\lambda \beta} + \eta^{\mu \alpha} \eta^{\lambda \beta} - \frac{1}{2} (\eta^{\mu \lambda} \eta^{\alpha \beta} + \eta^{\mu \lambda} \eta^{\alpha \beta}) \right] 
    -
\eta^{\mu \alpha} \bar{\mathfrak{g}}^{\lambda \beta} - \bar{\mathfrak{g}}^{\mu \alpha} \eta^{\lambda \beta} + \frac{1}{2} (\eta^{\mu \lambda} \bar{\mathfrak{g}}^{\alpha \beta} + \bar{\mathfrak{g}}^{\mu \lambda} \eta^{\alpha \beta})\right ] 
   \partial_{\lambda} h_{\alpha \beta} 
   \\ & 
   - \frac{\kappa}{2} B_{\mu}  
\left [ \eta^{\mu \alpha} \eta^{\lambda \beta}  - \frac{1}{2} \eta^{\mu \lambda} \eta^{\alpha \beta} \right] 
       \eta^{\sigma \rho}
        \left[ ( \partial_{\lambda} \bar{\mathfrak{g}}_{\rho \alpha} + \partial_{\alpha} \bar{\mathfrak{g}}_{\rho \lambda} - \partial_{\rho} \bar{\mathfrak{g}}_{\alpha \lambda} )
            h_{\sigma \beta}
+ 
         ( \partial_{\lambda} \bar{\mathfrak{g}}_{\rho \beta} + \partial_{\beta} \bar{\mathfrak{g}}_{\rho \lambda} - \partial_{\rho} \bar{\mathfrak{g}}_{\beta \lambda} )
        h_{\alpha\sigma}\right].
\end{split}
\end{equation}
In momentum space, the vertex can be written as: 
\begin{equation}\label{eq:VghB.eps}
    \vcenter{\hbox{\includegraphics[scale=0.6]{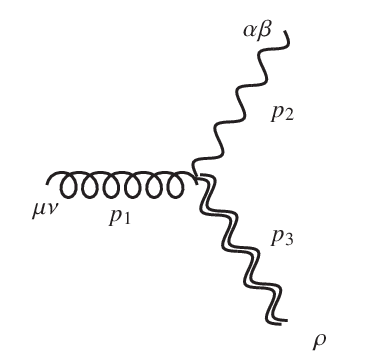}}}
    \quad
    \begin{aligned}
        \frac{\kappa}{4} & [p_{1}^{\nu } \eta^{\alpha \rho } \eta^{\beta \mu }+p_{1}^{\mu } \eta^{\alpha \rho }
   \eta^{\beta \nu }+p_{1}^{\nu } \eta^{\alpha \mu } \eta^{\beta \rho }+p_{1}^{\mu }
   \eta^{\alpha \nu } \eta^{\beta \rho }-p_{1}^{\beta } \eta^{\alpha \rho } \eta^{\mu \nu
   }-p_{1}^{\alpha } \eta^{\beta \rho } \eta^{\mu \nu }
   -p_{2}^{\rho } \eta^{\alpha \nu }
   \eta^{\beta \mu }\\ & +p_{2}^{\nu } \eta^{\alpha \rho } \eta^{\beta \mu }+p_{2}^{\alpha }
   \eta^{\beta \mu } \eta^{\nu \rho }-p_{2}^{\rho } \eta^{\alpha \mu } \eta^{\beta \nu
   }+p_{2}^{\mu } \eta^{\alpha \rho } \eta^{\beta \nu }+p_{2}^{\nu } \eta^{\alpha \mu }
   \eta^{\beta \rho }  +p_{2}^{\mu } \eta^{\alpha \nu } \eta^{\beta \rho } +p_{2}^{\rho }
   \eta^{\alpha \beta } \eta^{\mu \nu } \\ & -p_{2}^{\beta } \eta^{\alpha \rho } \eta^{\mu \nu
   }-p_{2}^{\alpha } \eta^{\beta \rho } \eta^{\mu \nu }-p_{2}^{\nu } \eta^{\alpha \beta }
   \eta^{\mu \rho }+p_{2}^{\beta } \eta^{\alpha \nu } \eta^{\mu \rho }+p_{2}^{\alpha }
   \eta^{\beta \nu } \eta^{\mu \rho }-p_{2}^{\mu } \eta^{\alpha \beta } \eta^{\nu \rho
   }+p_{2}^{\beta } \eta^{\alpha \mu } \eta^{\nu \rho }].
    \end{aligned}
\end{equation}

\subsection{Ghost vertex}
The ghost vertex is given by
\begin{equation}\label{eq:vertexghost}
    \vcenter{\hbox{\includegraphics[scale=0.6]{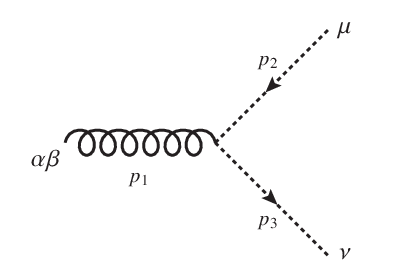}}}
    \qquad 
    \begin{aligned}
        i \frac{\kappa}{2}\big\{ & 
-{} p_3^2 \eta^{\alpha\beta} \eta^{\mu\nu}
+p_3^{\alpha}p_3^{\beta}\eta^{\mu\nu} + p_3^2 \eta^{\mu\alpha} \eta^{\nu\beta}
+{p}_1^{\nu} p_2^{\beta} \eta^{\mu\alpha}-{p}_1^{\mu} p_2^{\beta} \eta^{\nu\alpha}
                              \\ & -{p}_1\cdot p_3 \eta^{\mu\beta} \eta^{\nu\alpha}
+ 2  p_3^{\alpha} {p}_1^{\beta} \eta^{\mu\nu}
-{p}_1\cdot p_3 \eta^{\mu\nu}\eta^{\alpha\beta}
-{}
{p}_1^{\nu} p_2^{\beta} \eta^{\mu\alpha}
                              \\ &
+{p}_1^{\mu} p_2^{\alpha} \eta^{\nu\beta}
+{p}_1\cdot p_3 \eta^{\mu\beta} \eta^{\nu\alpha}
-{p}_1^{\alpha}\left({p}_1^{\mu}\eta^{\nu\beta}+{p}_1^{\nu}\eta^{\mu\beta}\right) 
\\&
+ {p}_1^{\mu} {p}_1^{\nu} \eta^{\alpha\beta} + {p}_1^2\eta^{\mu\alpha}\eta^{\nu\beta}
\big\}+ \mbox{ symmetrizations } \alpha\leftrightarrow\beta.
\end{aligned}
\end{equation}

\subsection{Gravity coupled to a scalar field}

The vertices with the matter field are presented below:
\allowdisplaybreaks
\begin{align}\label{eq:ppbpbp}
    & \vcenter{\hbox{\includegraphics[scale=0.7]{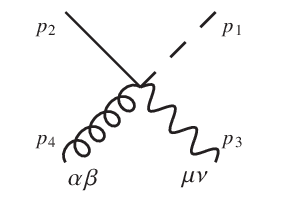}}}
    & & \begin{aligned}
        -i\frac{\kappa^2}{4  }& \big[-  p_{1}^{\beta} p_{2}^{\nu } \eta^{\alpha\mu }-  p_{1}^{\beta} p_{2}^{\mu } \eta^{\alpha\nu }-  p_{1}^{\alpha} p_{2}^{\nu } \eta^{\beta\mu }+  p_{1}^{\nu } (p_{2}^{\mu } \eta^{\alpha\beta}-p_{2}^{\beta} \eta^{\alpha\mu }-p_{2}^{\alpha} \eta^{\beta\mu })
                              \\ & -  p_{1}^{\alpha} p_{2}^{\mu } \eta^{\beta\nu }+  p_{1}^{\mu } (p_{2}^{\nu } \eta^{\alpha\beta}-p_{2}^{\beta} \eta^{\alpha\nu }-p_{2}^{\alpha} \eta^{\beta\nu })+  p_{2}^{\alpha} p_{1}^{\beta} \eta^{\mu \nu }+  p_{1}^{\alpha} p_{2}^{\beta} \eta^{\mu \nu }
                              \\ & -  p_{1}{} \cdot p_{2}{} \eta^{\alpha\beta} \eta^{\mu \nu }+  p_{1}{} \cdot p_{2}{} \eta^{\alpha\nu } \eta^{\beta\mu }+  p_{1}{} \cdot p_{2}{} \eta^{\alpha\mu } \eta^{\beta\nu }\big],
    \end{aligned}   
 \\   &  
    \vcenter{\hbox{\includegraphics[scale=0.6]{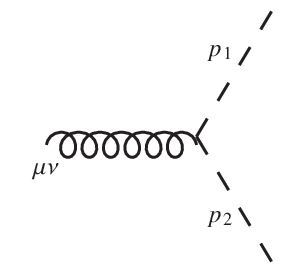}}}
   && 
    i \frac{\kappa}{2}\left ( \eta_{\mu \nu} {p}_{1} \cdot p_{2}   - {p}_{1 \mu} {p}_{2 \nu} - {p}_{1 \nu} {p}_{2 \mu}\right ),
  \\  & 
    \vcenter{\hbox{\includegraphics[scale=0.7]{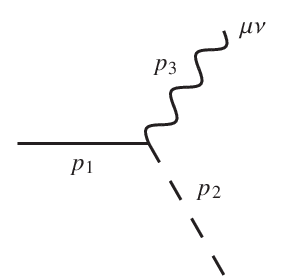}}} 
      && i \frac{\kappa}{2} \left[\eta^{\mu \nu } 
   {{p}}_1\cdot
   {p_2}
   -
    {p_2}^{\mu } {{p}}_1^{\nu
   }-   {{p}}_1^{\mu }
   {p_2}^{\nu }
\right],
\\\label{eq:quartic_vertex}
    &    \vcenter{\hbox{\includegraphics[scale=0.75]{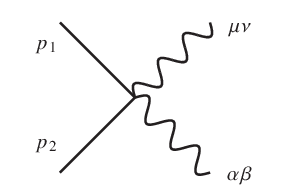}}}
    &&
        \begin{aligned}   -i \frac{\kappa^{2}}{4}
        & \big[ 
 p_2^{\mu }
   p_1^{\nu } \eta^{\alpha \beta }+
    p_1^{\mu }
   p_2^{\nu } \eta^{\alpha \beta }-
    p_2^{\beta }
   p_1^{\nu } \eta^{\alpha \mu }-
    p_1^{\beta }
        p_2^{\nu } \eta^{\alpha \mu }  -
    p_2^{\beta }
   p_1^{\mu } \eta^{\alpha \nu }-
    p_1^{\beta }
            p_2^{\mu } \eta^{\alpha \nu } \\ & -
    p_2^{\alpha }
   p_1^{\nu } \eta^{\beta \mu }-
    p_1^{\alpha }
        p_2^{\nu } \eta^{\beta \mu } -
    p_2^{\alpha }
   p_1^{\mu } \eta^{\beta \nu }-
    p_1^{\alpha }
   p_2^{\mu } \eta^{\beta \nu }+
    p_2^{\alpha }
   p_1^{\beta } \eta^{\mu \nu }+
    p_1^{\alpha }
        p_2^{\beta } \eta^{\mu \nu }\\ & +
    \eta^{\alpha \nu } \eta^{\beta \mu }
   p_1\cdot
   p_2+
   \eta^{\alpha \mu } \eta^{\beta \nu }
   p_1\cdot
   p_2-
   \eta^{\alpha \beta } \eta^{\mu \nu }
   p_1\cdot
   p_2\big].
    \end{aligned}
\end{align}

\section{One-Loop counter-terms} \label{sec:OneLoop}
Here, we compute the diagrams in Figs.~\ref{fig:2point}, \ref{fig:3point} and \ref{fig:scalar_scalar} in the Landau background gauge
using the Feynman rules derived in the Appendix~\ref{sec:FR}. We will use dimensional regularization with $ D = 4 -2 \epsilon $, where $D$ is the spacetime dimension.
With these results, we derive the corresponding one-loop counter-term Lagrangians.

\subsection{$2$-point diagrams}
Here, we present the divergent part of the diagrams shown in Fig.~\ref{fig:2point}. We introduce the following tensor basis:
\begin{subequations} \label{eq:basis}\allowdisplaybreaks
\begin{align}
    \mathcal{T}^{\mu\nu\alpha\beta}_{(1)} ={}&        k^{\mu} k^{\alpha} k^{\nu} k^{\beta},\\
    \mathcal{T}^{\mu\nu\alpha\beta}_{(2)} ={}& \eta^{\mu\nu}\eta^{\alpha\beta},\\
    \mathcal{T}^{\mu\nu\alpha\beta}_{(3)} ={}& \eta^{\mu\beta} \eta^{\alpha\nu}+\eta^{\mu\alpha}\eta^{\nu\beta},\\
    \mathcal{T}^{\mu\nu\alpha\beta}_{(4)} ={}& k^{\alpha} k^{\beta} \eta^{\mu\nu}+k^{\mu}k^{\nu} \eta^{\alpha\beta},\\
    \mathcal{T}^{\mu\nu\alpha\beta}_{(5)} ={}& k^{\nu} k^{\beta}\eta^{\mu\alpha}+k^{\alpha} k^{\nu} \eta^{\mu\beta}+k^{\mu}k^{\beta} \eta^{\alpha\nu}+k^{\mu} k^{\alpha} \eta^{\nu\beta}.
\end{align}
\end{subequations}

\subsubsection{Graviton self-energy}
First, we compute the graviton loop, which corresponds to the diagrams depicted in Fig.~\ref{fig:2point}(a, b, c, d). 
These diagrams lead to the following counter-terms
\begin{equation}\label{eq:VH_grav}
    \mathcal{L}_{\text{2(a+b+c+d)}} =  \frac{\sqrt{\bar{g}} }{16 \pi^{2} \epsilon} \frac{7}{40} \left (   \bar{R}^{2}   + 2   \bar{R}_{\mu \nu}^{2}\right ).
\end{equation}

Next, we consider the scalar loop diagram shown in Fig.~\ref{fig:2point}(e).
In the basis \eqref{eq:basis}, we obtain that the scalar loop diagram is equal to
\begin{equation}\label{eq:scalarloopres}
    \frac{\kappa ^2}{16 \pi^{2} \epsilon }
    \left( \frac{1}{60}\mathcal{T}_1^{\mu \nu \alpha \beta}  +  \frac{1}{80}\mathcal{T}_2^{\mu \nu \alpha \beta}      k^4 + \frac{1}{480}\mathcal{T}_3^{\mu \nu \alpha \beta}   k^4 -\frac{1}{80}\mathcal{T}_4^{\mu \nu \alpha \beta}   k^2-\frac{1}{480}\mathcal{T}_5^{\mu \nu \alpha \beta}  k^2.
   \right)
\end{equation}
This result leads to the following counterterms
\begin{equation}\label{eq:VH_scalar_lag}
    \mathcal{L}_{\text{2e}} =  \frac{\sqrt{\bar{g}} }{16 \pi^{2} \epsilon} \frac{1}{240} \left (  \bar{R}^{2}+ 2\bar{R}_{\mu \nu}^{2} \right ).
\end{equation}

One can find the graviton self-energy in the Landau gauge by summing the diagrams in Fig.~\ref{fig:2point}(a, b, c, d, e).
The divergent part of the graviton self-energy leads to the counterterm Lagrangian:
\begin{equation}\label{eq:total2}
    \mathcal{L}_{\text{2(a+b+c+d+e)}} =  \frac{\sqrt{\bar{g}} }{16 \pi^{2} \epsilon} \frac{43}{240}\left (
      \bar{R}^{2}
 +
     2 \bar{R}_{\mu \nu}^{2}
 \right ).
\end{equation}

\subsubsection{Scalar self-energy $ \Sigma $}
The scalar self-energy $ \Sigma $ is shown in Fig.~\ref{fig:2point}(f).
The divergent part of this diagram vanishes in the Landau gauge: 
\begin{equation}\label{eq:sigma2}
    \Sigma^{\text{div}} =0.
\end{equation}

\begin{figure}[ht]
    \includegraphics[width=0.45\textwidth]{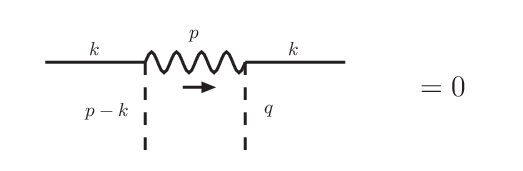}
    \caption{
Diagrammatic identity between the background graviton-scalar-scalar vertex and the graviton propagator in Landau gauge. The arrow indicates the momentum flow. All external lines are amputated.
    }\label{fig:landause}
\end{figure}

Since the identity shown in Fig.~\ref{fig:landause} holds, in any spacetime dimension $D$ in the Landau gauge, the scalar self-energy must vanish in any spacetime dimension $ D$ as well.

\subsection{$3$-point function}\label{sec:3point}

The one-loop diagrams that contributes to the $3$-point function $ \bar{\mathfrak{g}}  (k) {\bar{\phi}} (k_1) {\bar{\phi}} (k_{2} )$ are shown in Fig.~\ref{fig:3point}. In order to simplify our results, we will introduce the tensor basis:
\begin{subequations}\label{eq:tensorbasis2}\allowdisplaybreaks
\begin{align}
    \mathcal{T}^{\mu \nu}_{\bar{R} } (k,k_{1} , k_{2} )={}&
   -2 k_{1} \cdot k_{2 } ( k^{\mu} k^{\nu} - k^{2} \eta^{\mu \nu} ), \\\nonumber
    \mathcal{T}^{\mu \nu}_{\bar{R}_{ \alpha \beta } }(k,k_{1} , k_{2} ) ={}&
\left({k}\cdot {k}_1\right) \left({k}\cdot {k}_2\right)
\eta^{\mu \nu } + \frac{k^{2}}{2} ( {k}_2{}^{\mu } {k}_1{}^{\nu
   }+  {k}_1{}^{\mu } {k}_2{}^{\nu })\\ & -\frac{1}{2} \big[
   {k}_2{}^{\mu } {k}^{\nu } \left({k}\cdot
   {k}_1\right)+ {k}^{\mu } {k}_2{}^{\nu }
   \left({k}\cdot {k}_1\right)+{k}_1{}^{\mu }
   {k}^{\nu } \left({k}\cdot {k}_2\right)+
   {k}^{\mu } {k}_1{}^{\nu } \left({k}\cdot {k}_2\right)
\big]
   .
\end{align}
\end{subequations}
The above tensor structures corresponds, in momentum space, to the invariants: $ \bar{R} \partial_{\mu} \bar{\phi} (k_{1} ) \partial^{\mu} \bar{\phi} ( k_{2} )$ and $ \bar{R}_{\mu \nu} (k) \partial^{\mu} \bar{\phi} ( k_{1} ) \partial^{\nu} \bar{\phi} (k_{2} )$. 

Using this basis, one can show that the sum of the divergent part of the diagrams Fig.~\ref{fig:3point}(a,b) can be written as 
    \begin{align}\label{eq:3pAa}
&
\frac{4}{3} \kappa ^3  \left [ \frac{1}{4} \mathcal{T}_{ \bar{R} }^{\mu \nu} (k, k_{1} , k_{2} )- \mathcal{T}_{\bar{R}_{\alpha \beta }}^{\mu \nu} (k, k_{1} , k_{2} )\right ].
    \end{align}

Now, we consider the divergent part of the diagrams in Fig.~\ref{fig:3point}(c, d, e, f). Summing all these contributions, we obtain a transverse result that can be written in terms of the basis \eqref{eq:tensorbasis2}:
\begin{equation}\label{eq:sum3def}
    \frac{ \kappa^{3} }{2}  \mathcal{T}^{\mu \nu}_{\bar{R}_{\alpha \beta } }
( k_{1} + k_{2} , k_{1} , k_{2} ).
\end{equation}

The total result, using Eqs.~\eqref{eq:3pAa} and \eqref{eq:sum3def}, leads to the counterterms:
\begin{equation}\label{eq:3pointCT}
    \mathcal{L}_{\text{3P}}  = \kappa^{2} \frac{ \sqrt{\bar{g} } }{16 \pi^{2} \epsilon}  \left [ \frac{1}{3}  \bar{R} \partial_{\mu} \bar{\phi} \partial^{\mu} \bar{\phi}
        -\frac{5}{6} 
        \bar{R}_{\mu \nu} \partial^{\mu} \bar{\phi} \partial^{\nu} \bar{\phi}
    \right ].
\end{equation}

\subsection{$4$-point scalar function}
Next, let us consider the divergent part of the $4$-point scalar function. Note that permutations are implied.

The diagram shown in Fig.~\ref{fig:scalar_scalar}(a) (summing all the permutations) leads to the following divergent contribution:
\begin{equation}\label{eq:4point}
    2  \kappa ^4 \left[\left({k_1}\cdot
   {k_4}\right)
   \left({k_2}\cdot
   {k_3}\right)+\left({k_1}\cdot {k_3}\right)
   \left({k_2}\cdot
   {k_4}\right)+\left({k_1}\cdot {k_2}\right)
   \left({k_3}\cdot
   {k_4}\right)\right].
\end{equation}

Now, one can compute the divergent part of the triangle and box diagrams shown in Fig.~\ref{fig:scalar_scalar}(b, c) to show that both vanishes. On the other hand, by power counting, it is easy to see that both diagrams are finite in the Landau gauge. 

The finiteness of diagram~\ref{fig:scalar_scalar}(c) comes from a generalization of the identity shown in Fig.~\ref{fig:landause}. In the Landau gauge, we have that 
    \begin{equation}\label{eq:id2}
        \vcenter{\hbox{\includegraphics[scale=0.75]{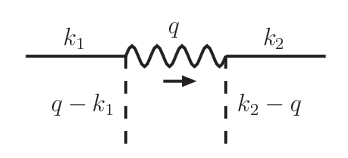}}} =i \kappa^{2} \frac{q_{\mu} q_{\nu}}{q^4} 
        \left [ -\mathcal{T}_{\bar{R} }^{\mu \nu} (k_{2} - k_{1} , k_{1} , k_{2} )+ 2\mathcal{T}_{\bar{R}_{\alpha \beta }}^{\mu \nu} (k_{2} - k_{1} , k_{1} , k_{2} )\right ],
\end{equation}
which is the top-half of the graph (when $k_1=k_2=k$, Eq.~\eqref{eq:id2} reduces to the identity in Fig.~\ref{fig:landause}). Joining it with the bottom-half and using above identity for both parts, we obtain that the loop integrand has the momentum dependence 
\begin{equation}\label{eq:boxdiv}
    \underbrace{p^{D}}_{\text{loop}} \underbrace{\frac{p_{\mu} p_{\nu}}{p^4}}_{\text{top-half}} \underbrace{\frac{1}{p^{2}}}_{\text{scalar}} \underbrace{\frac{p_{\alpha} p_{\beta}}{p^{4}}}_{\text{bottom-half}} \underbrace{\frac{1}{p^{2}}}_{\text{scalar}}  \sim  \frac{I_{\mu \nu \alpha \beta}}{p^{4}} ,
\end{equation}
where $ D =4$ is the dimension of the spacetime. 
By power counting, we find that the box diagram is finite. An analogous argument applies to the triangle diagram in Fig.~\ref{fig:scalar_scalar}(b), but now it relies on the observation that the vertex defined in Eq.~\eqref{eq:quartic_vertex} depends solely on external momenta.

Thus, the counterterm Lagrangian, in the Landau gauge, induced by the $4$-point function is given by
\begin{equation}\label{eq:CT_4point}
    \begin{split}
        \mathcal{L}_{\text{4P}} =\frac{1}{4}  \kappa^{4}
        \frac{\sqrt{\bar{g}}}{16 \pi^{2} \epsilon}   {( \bar{g}^{\mu \nu}\partial_{\mu} \bar{\phi}  \partial_\nu \bar{\phi} )}^{2}
\end{split}.
\end{equation}

Adding all contribution [Eqs.~\eqref{eq:total2}, \eqref{eq:sigma2}, \eqref{eq:3pointCT} and \eqref{eq:CT_4point}], we obtain the off-shell background counter-term Lagrangian \eqref{eq:310}.

\bibliography{LandauGauge.bib}
\end{document}